\documentclass[a4paper,11pt]{article}
\pdfoutput=1 

\usepackage{jcappub} 
\usepackage[T1]{fontenc} 

\title{\boldmath Exact cosmological black hole solutions in Scalar Tensor Vector Gravity}

\author[a]{D. P\'erez,\note{Corresponding author.}}
\author[a,b]{G. E. Romero}

\affiliation[a]{Instituto Argentino de Radioastronom{\'\i}a \\
 (CCT - La Plata, CONICET; CICPBA), Camino Gral Belgrano Km 40 C.C.5, 1894 Villa Elisa, Buenos Aires, Argentina\\}
\affiliation[b]{Facultad de Ciencias Astron\'omicas y Geof{\'\i}sicas, UNLP, Paseo del Bosque s/n CP, 1900 La Plata, Buenos Aires, Argentina}

\emailAdd{danielaperez@iar.unlp.edu.ar}
\emailAdd{romero@iar.unlp.edu.ar}

\abstract{We find an exact solution of Scalar-Tensor-Vector Gravity field equations that represents a black hole embedded in an expanding universe. This is the first solution of the kind found in the theory. We analyze the properties of the apparent horizons as well as the essential singularities of  the metric, and compare it with the McVittie spacetime of General Relativity. Depending on the cosmological model adopted and the value of the free parameter $\alpha $ of the theory, the solution describes a cosmological black hole, an inhomogeneity in an expanding universe, or a naked singularity.  We use the latter result to set further constraints  on the free parameters of the theory.}

\begin{document}
\maketitle
\flushbottom

\section{Introduction}
\label{sec:intro}

``Probably the most beautiful of all existing theories''. These words by Landau and Lifschitz \cite{lan+71} reflect the pleasant aesthetic experience induced on many of us by General Relativity (GR). The theory not only excels in simplicity, symmetry,  unification strength, and fundamentality \cite{rom18}, but also has an outstanding predictive and explanatory power. Though Einstein himself remarked GR charm \cite{ein15b}, he was quite aware that it was not the ultimate theory of gravitation. He struggled the last decades of his life searching for suitable generalizations of the theory that could accommodate electrodynamics and also include quantum effects.

Besides the inherent deficiencies in the theory, such as the problem of spacetime singularities, GR models do not succeed in reproducing rotation curves of nearby galaxies, mass profiles of galaxies clusters, some gravitational lensing effects, and cosmological data. A possible solution to these problems consist in modifying the right hand side of Einstein equations: a term with a cosmological constant is added and the existence of dark matter is postulated. From an ontological point of view, this approach is quite costly since we are assuming the existence of entities of unknown nature whose properties have never been measured to date \cite{apr+12,lux+13,agn+14}.

We can follow a different strategy to explain the astronomical data: modify the theory of gravitation. This is the case of Scalar-Tensor-Vector Gravity (STVG), also dubbed MOdified gravity (MOG) \cite{mof06}. In STVG, the effects of gravity are not only represented by a metric tensor field but also by a scalar and a vector field. Specifically, the universal constant $G$ along with the mass $\tilde{\mu}$ of the vector field are the dynamical scalar fields of the theory. When gravity is weak, the equations of the theory reduce to a modified acceleration law characterized by: 1) an enhanced Newtonian constant $G = G_{\rm N} \left(1 + \alpha\right)$, and 2) at certain scales, a repulsive Yukawa force term that counteracts the augmented Newtonian acceleration law, in such a way that in the Solar System GR is recovered.  The first of the features  mentioned above allows to reproduce the rotation curves of many galaxies \cite{bro+06a, mof+13,mof+15}, the dynamics of galactic clusters \cite{bro+06b, bro+07, mof+14}and cosmological observations \cite{mof+07, mof15c}, without dark matter\footnote{The recent detection of a neutron star merger in gravitational waves \cite{abb+17a} (GW170817), and the subsequent observation of the electromagnetic counterpart GRB 170814A \cite{abb+17b,abb+17c} has been used to show that a large class of alternative theories of gravitation, for instance those in which photons suffer an additional Shapiro time delay,  must be discarded \cite{ezq+17, par+18}. As demonstrated by Green and collaborators \cite{gre+18}, STVG survives such stringent test: both gravitational and electromagnetic travel on null geodesics in the theory.}.

We can classify the known solutions of the field equations of STVG in two main groups.  On the one hand, vacuum and non-vacuum solutions for a given distribution of matter where the spacetime metric is asymptotically flat. This is the case of the Schwarzschild and Kerr STVG black holes\footnote{Different aspects of the STVG black hole solutions have been extensively studied in the literature: accretion disks around Schwarzschild and Kerr STVG black holes \cite{per+17}, shadows cast by near-extremal Kerr STVG black holes \cite{guo+18}, black hole superradiance in STVG \cite{won+18}, quasinormal modes of Schwarzschild STVG black holes \cite{man+18}, the process of acceleration and collimation of relativistic jets in Kerr STVG black holes \cite{lop+17b}, dynamics of neutral and charged particles around a Schwarzschild
STVG black hole immersed in a weak magnetic field \cite{hus+15}, among others.} found by Moffat \cite{mof15a}, and neutron star models constructed by Lopez Armengol and Romero \cite{lop+17a}. On the other hand, there are cosmological solutions such as the ones derived by Roshan \cite{ros15} and Jamali and collaborators \cite{jam+18}.  Until now, a third class of solutions remains unexplored in the theory: metrics that represent an inhomogeneity in an expanding universe.

In General Relativity, McVittie \cite{mcv33} was the first to obtain an exact solution of Einstein field equations that corresponds to a central inhomogeneity embedded in a Friedmann-Lema\^{i}tre-Robertson-Walker (FLRW) background. The McVittie metric and its generalization have been widely studied through the years (see for instance the works by Faraoni and Jacques \cite{far+07} and Carrera and Giulini \cite{car+10}).  The investigation of such solutions has transcended GR to encompass alternative theories of gravitation \cite{tre+18}. The results of the studies of inhomogeneous spacetimes have direct astrophysical implications: a cosmological force acting on large scales can modify the structure of galaxies and clusters of galaxies, and inhibit accretion processes. The effects of the cosmological expansion, thus, need to be taken into account when modeling the evolution of structure in the universe.

In this work we present exact solutions of STVG that represent an inhomogeneity in an expanding spacetime, and analyze the corresponding properties. We distinguish the metrics that represent cosmological black holes and we compare them with the corresponding solutions in GR.

The paper is organized as follows. We provide a brief introduction to STVG in Section \ref{sec2}. Next, we introduce a solution to the field equations of the theory that represents an inhomogeneity in an expanding universe. In Section \ref{sec4}, we analyze the properties of the metric: singularities and apparent horizons, and in Section \ref{sec5} we offer a discussion of the results obtained. The last section of the paper is devoted to the conclusions.

\section{\label{sec2}STVG gravity}

\subsection{STVG action and field equations}

The action \footnote{As suggested by Moffat and Rahvar \cite{mof+13} and  Moffat and Toth \cite{mof+09}, we dismiss the scalar field $\omega$, and we treat it as a constant, $\omega =1$.} in STVG theory is \cite{mof06}:
\begin{equation}\label{action}
S = S_{\rm GR} + S_{\phi} + S_{\rm S} + S_{\rm M},
\end{equation}
where
\begin{eqnarray}
S_{\rm GR} & = & \frac{1}{16 \pi} \int d^{4}x \sqrt{-g} \frac{1}{G} R ,\\
S_{\phi} & = & - \int d^{4}x \sqrt{-g}  \left(\frac{1}{4} B^{\mu\nu}B_{\mu\nu} - \frac{1}{2}\tilde{\mu}^2 \phi^{\mu} \phi_{\mu}\right),\\
S_{\rm S} & = & \int d^{4}x \sqrt{-g} \frac{1}{G^{3}} \left(\frac{1}{2} g^{\mu \nu} \nabla_{\mu}G\nabla_{\nu}G-V(G)\right)\\
& + &  \int d^{4}x \frac{1}{\tilde{\mu}^2 G} \left(\frac{1}{2} g^{\mu \nu} \nabla_{\mu}\tilde{\mu}\nabla_{\nu}\tilde{\mu}-V(\tilde{\mu})\right).
\end{eqnarray}
Here, $g_{\mu \nu}$ is the spacetime metric, $R$ denotes the Ricci scalar, and $\nabla_{\mu}$ is the covariant derivative; $\phi^{\mu}$ stands for a Proca-type massive vector field, $\tilde{\mu}$ is
its mass, and $B_{\mu \nu} = \partial_{\mu} \phi_{\nu} - \partial_{\nu}\phi_{\mu}$. The scalar fields $G(x)$ and $\tilde{\mu}(x)$ vary in space and time, and $V(G)$, and $V(\tilde{\mu})$ 
are the corresponding potentials. We adopt the metric signature $\eta_{\mu \nu} = {\rm diag}(-1,+1,+1,+1)$. 
The term $S_{\rm M}$ in the action refers to possible matter sources.

The full energy-momentum tensor for the gravitational sources is:
\begin{equation}
T_{\mu \nu} = T^{\rm M}_{\mu \nu} + T^{\phi}_{\mu \nu} + T^{\rm S}_{\mu \nu},
\end{equation} 
where
\begin{eqnarray}
T^{\rm M}_{\mu \nu} & = & -\frac{2}{\sqrt{-g}} \frac{\delta S_{\rm M}}{\delta g^{\mu \nu}},\\
T^{\phi}_{\mu \nu} & = & -\frac{2}{\sqrt{-g}} \frac{\delta S_{\phi}}{\delta g^{\mu \nu}},\\
T^{\rm S}_{\mu \nu} & = & -\frac{2}{\sqrt{-g}} \frac{\delta S_{\rm S}}{\delta g^{\mu \nu}}.
\end{eqnarray}
Following the notation introduced above, $T^{\rm M}_{\mu \nu}$ denotes the ordinary matter energy-momentum and $T^{\rm S}_{\mu \nu}$ the scalar contributions to the energy-momentum tensor; $T^{\phi}_{\mu \nu}$ stands for the energy-momentum tensor\footnote{ Moffat \cite{mof15a} set the potential $V(\phi)$ equal to zero in the definition of $T^{\phi}_{\mu \nu}$ given in \cite{mof06}.} of the field $\phi^{\mu}$:
\begin{equation}
T^{\phi}_{\mu \nu} = - \frac{1}{4}\left({B_{\mu}}^{\alpha} B_{\nu \alpha} - \frac{1}{4} g_{\mu \nu} B^{\alpha \beta} B_{\alpha \beta}\right).
\end{equation}

 The equation of motion for a test particle in coordinates $x^{\mu}$ is given by
\begin{equation}\label{eq-motion}
\left(\frac{d^{2}x^{\mu}}{d\tau^2} + \Gamma^{\mu}_{\alpha \beta} \frac{dx^{\alpha}}{d\tau}\frac{dx^{\beta}}{d\tau} \right) =\frac{q}{m} {B^{\mu}}_{\nu} \frac{dx^{\nu}}{d\tau},
\end{equation}
where $\tau$ represents the particle proper time, and $q$ is the coupling constant with the vector field.

Moffat \cite{mof15a}  postulates that the gravitational source charge $q$ of the vector field $\phi^{\mu}$ is proportional to the mass of the source particle,  
\begin{equation}\label{q}
q = \pm \sqrt{\alpha G_{N}} m.
\end{equation}
Here, $G_{\rm N}$ denotes Newton's gravitational constant, and $\alpha$ is a free dimensionless parameter. The positive value for the root is chosen ($q > 0$) to maintain a repulsive, gravitational Yukawa-like force when the mass parameter $\tilde{\mu}$ is non-zero. We see, then, that in STVG the nature of the gravitational field has been modified with respect to GR in two ways: there is an enhanced gravitational constant $G = G_{\rm N} \left(1 + \alpha \right)$, and a vector field $\phi^{\mu}$ that exerts a gravitational Lorentz-type force on any material object through Eq. \eqref{eq-motion}.


\section{\label{sec3}Solution of STVG field equations}

\subsection{Derivation of the metric}

In order to derive the spacetime metric that represents an inhomogeneity in an expanding universe, we make the following assumptions:
\begin{itemize}
 \item The energy-momentum tensor has two components $T_{\mu \nu} = T^{\mathrm M}_{\mu \nu} + T^{\phi}_{\mu \nu}$, where $T^{\mathrm M}_{\mu \nu}$ stands for the energy-momentum of the cosmological fluid, and $T^{\phi}_{\mu \nu}$ is the energy-momentum tensor for the vector field $\phi^{\mu}$:
 \begin{equation}
  T^{\mathrm M}_{\mu \nu}  = \left(\rho + \frac{p}{c^{2}}\right) u_{\mu} u_{\nu} + p g_{\mu \nu}.\label{cosmo-fluid}\\
\end{equation}
Here,  $\rho$ and $p$ are the density and pressure of the cosmological fluid, respectively, and $u^{\mu}$ is the four-velocity of the fluid that in a comoving coordinate system has the form:
\begin{equation}
\left[u^{\mu}\right] =   \left(\frac{c}{\sqrt{- g_{00}}}, 0, 0, 0\right).
\end{equation}

\item Since the effects of the mass of the vector field $\tilde{\mu}$ manifest on kiloparsec scales from the source, it is neglected when solving the field equations for compact objects such as black holes \cite{mof15a}.

\item $G$ is a constant that depends on the parameter $\alpha$ \cite{mof06}:
\begin{equation}\label{grav-const}
G = G_{\rm N} \left(1 + \alpha \right).
\end{equation} 
\end{itemize}

Given these hypotheses, the action \eqref{action} takes the form,
\begin{equation}\label{action1}
S =   \int d^{4}x \sqrt{-g} \left( \frac{R}{16 \pi G} -\frac{1}{4}B^{\mu\nu} B_{\mu\nu}\right).
\end{equation}
Variation of the latter expression with respect to $g_{\mu \nu}$ yields the STVG field equations:
\begin{equation}\label{fe1}
G_{\mu \nu} = 8 \pi G \left(T^{\mathrm M}_{\mu \nu} +T^{\phi}_{\mu \nu}\right),
\end{equation}
where $G_{\mu \nu}$ is the Einstein tensor.
If we vary the action \eqref{action1} with respect to the vector field $\phi^{\mu}$, we obtain the dynamical equation for this field:
\begin{equation}\label{tb}
\nabla_{\nu}B^{\mu \nu} = 0,
\end{equation}
and
\begin{equation}\label{tb1}
\nabla_{\sigma}B_{\mu \nu} +\nabla_{\mu}B_{\nu \sigma}+\nabla_{\nu} B_{\sigma \nu}= 0.  
\end{equation}

We propose the following metric ansatz\footnote{In what follows we work with geometrized units $G = c =1$.}:
\begin{equation}\label{line-ele}
ds^{2} = - {A(t,x)}^{2} dt^{2}  + {B(t,x)}^{2} \left({dx}^2 + x^{2} {d\theta}^{2} + x^{2} {\sin{\theta}}^{2} {d\phi}^{2}\right),
\end{equation}
where $(t,x,\theta,\phi)$ are isotropic coordinates. Since the off-diagonal elements of the energy-momentum tensor $T_{\mu \nu} = T^{\mathrm M}_{\mu \nu} + T^{\phi}_{\mu \nu}$ are zero, the $G_{\mathrm{tx}}$ component of the Einstein tensor yields:
\begin{equation}\label{gtx}
G_{\mathrm{tx}} = 0.  
\end{equation}
If we calculate $G_{\mathrm{tx}}$ and substitute in Eq. \eqref{gtx}, after some algebra we get:
\begin{equation}\label{A}
 A(t,x) = \frac{f(x)}{2}\frac{\dot{B}(t,x)}{B(t,x)}.
\end{equation}

If we now compare the line element given by Eq. \eqref{line-ele} with the line element for a Schwarzschild STVG black hole\footnote{The coordinate transformation between Schwarzschild coordinates $(t,r,\theta,\phi)$ and isotropic coordinates $(t,x,\theta,\phi)$  is:
\begin{equation}
r = x   \left[\left(1 + \frac{M}{2 x}\right)^{2} - \frac{Q^{2}}{4 x^{2}}\right].
\end{equation}} in isotropic coordinates:
\begin{eqnarray}\label{schw-iso}
ds^{2} & = & - \frac{\left(1 - \frac{M^{2}}{4 x^{2}} + \frac{Q^{2}}{4 x^{2}}\right)^{2}}{\left[\left(1 + \frac{M}{2 x}\right)^{2} - \frac{Q^{2}}{4 x^{2}}\right]^{2}}  {dt}^{2}\nonumber \\
& + &  \left[\left(1 + \frac{M}{2 x}\right)^{2} - \frac{Q^{2}}{4 x^{2}}\right]^{2} \left({dx}^{2} + x^{2} {d\Omega}^{2}\right),
\end{eqnarray}
where ${d\Omega}^{2} = {d\theta}^{2} + {\sin{\theta}}^{2} {d\phi}^{2}$, a possible form for $B(t,x)$ is:
\begin{equation}\label{B}
B(t,x) = \left[k(t,x) + \frac{l(t)}{x}\right]^{2} - \frac{h(t)}{x^{2}}.
\end{equation}
The functions $l(t)$ and $h(t)$ are related to the gravitational mass $M$ and gravitational charge $Q$ of the source. Because we assume that both $M$ and $Q$ are not distributed in space but are concentrated in the singularity, $l(t)$ and $h(t)$ depend only on the time coordinate. By substituting Eq. \eqref{B} into Eq. \eqref{A}, we derive an expression for $A(t,x)$:
\begin{equation}
A(t,x) = f(t)\frac{\frac{\dot{k}}{k} + \frac{l \dot{l}}{k^{2} x^{2}} +\frac{\left(\dot{l} k + l \dot{k}\right)}{x k^{2}} - \frac{\dot{h}}{2 k^{2} x^{2}}  }{\left[\left(1 + \frac{l}{x k}\right)^{2} - \frac{h}{k^{2} x^{2}}\right]}.  
\end{equation}

To determine the specific form of the functions $f(t)$, $k(t,x)$, $l(t)$, and $h(t)$, we require that $A(t,x) \rightarrow \sqrt{g_{\mathrm{tt}}}$ in \eqref{schw-iso} for $t = \mathrm{const}$. This yields:
\begin{eqnarray}
f \frac{\dot{k}}{k} & = & 1 \;\;\;  \Rightarrow \;\;\;  f\dot{k} = k, \label{c1}\\
f \frac{\left(\dot{l} k + l \dot{k}\right)}{x k^{2}} & = & 0\;\;\; \Rightarrow \;\;\; f \dot{l}= - l, \label{c2}\\
f  \frac{l \dot{l}}{k^{2} x^{2}} & = & - \frac{l^{2}}{x^{2} k^{2}}, \label{c3}\\
-  \frac{f \dot{h}}{2 k^{2} x^{2}} & = &  \frac{h}{k^{2} x^{2}} \;\;\; \Rightarrow \;\;\; \dot{h} f = -2 h.\label{c4}
\end{eqnarray}

In the limit $l(t) \rightarrow 0$, $h(t) \rightarrow 0$ (the gravitational mass and charge tend to zero), the line element should be that of a FLRW model (for simplicity we assume the spatial curvature $\kappa = 0$):
\begin{equation}
ds^{2} = - dt^{2} + {a(t)}^{2} \left(dx^{2} + x^{2} {d\Omega}^{2}\right).  
\end{equation} 
Thus, the function $k$ depends only on the temporal coordinate, and from \eqref{c1}:
\begin{equation}
f = \frac{k}{\dot{k}} = \frac{\tilde{a}(t)}{\dot{\tilde{a}}(t)}.
\end{equation}
Substituting the later expression into Eqs. \eqref{c2} and \eqref{c4} yields:
\begin{eqnarray}
  \dot{l} \frac{\tilde{a}}{\dot{\tilde{a}}} & = &- l \;\;\; \Rightarrow \;\;\; l =\frac{M}{\tilde{a}},\\
  \dot{h}\frac{\tilde{a}}{\dot{\tilde{a}}} & = & -2 h \;\;\; \Rightarrow \;\;\; h= \frac{Q^{2}}{{\tilde{a}}^{2}}.
\end{eqnarray}
The integration constants $M$ and $Q$ are the gravitational mass and gravitational charge of the central inhomogeneity, respectively, while $\tilde{a}$ is associated with the scale factor $a(t)$ of the cosmological model as $\tilde{a}(t) = \sqrt{a(t)}$. 

Finally, by replacing $f(t)$, $k(t)$, $l(t)$, and $h(t)$ into expressions \eqref{A} and \eqref{B}, the metric \eqref{line-ele} takes the form:
\begin{eqnarray}
ds^{2} & = & - c^{2} \frac{\left[1- \frac{G \left(G M^{2} - Q^{2}\right)}{4 c^{4} a^{2} x^{2}}\right]^{2}}{\left[\left(1 +\frac{G M}{2 c^{2} x a}\right)^{2} - \frac{G Q^{2}}{4 c^{4} a^{2} x^{2}}\right]^{2}} dt^{2} \label{metric}\\
& +& {a(t)}^{2} \left[\left(1 +\frac{G M}{2 c^{2} x a}\right)^{2} - \frac{G Q^{2}}{4 c^{4} a^{2} x^{2}}\right]^{2} \left(dx^{2} + x^{2} {d\Omega}^{2}\right),\nonumber
\end{eqnarray}
where the corresponding constants have been adequately restored.

In the next subsection, we prove that the metric here obtained does indeed satisfy the field equations of the theory.

\subsection{Correctness of the metric}

The first step in order to show that metric \eqref{metric} satisfies the field equations of STVG given by \eqref{fe1}, \eqref{tb}, and \eqref{tb1}, is to compute the Einstein tensor. The non-zero components of $G^{\mu}_{\nu}$ are:
\begin{eqnarray}
{G^{t}}_{t} & = &  -a_{0} - 3 \left(\frac{\dot{a}}{a}\right)^{2} = \kappa \left(- \rho c^{2} + \frac{B^{t x} B_{t x}}{8 \pi} \right),\label{gtt}\\
 {G^{x}}_{x} & = & - a_{0} -  \left(\frac{\dot{a}}{a}\right)^{2} a_{1} - 2 \frac{\ddot{a}}{a} a_{2} =  \kappa \left( p + \frac{B^{t x} B_{t x}}{8 \pi} \right)\label{grr}\\
 {G^{\theta}}_{\theta} & = &   a_{0} -  \left(\frac{\dot{a}}{a}\right)^{2} a_{1} - 2 \frac{\ddot{a}}{a} a_{2} =  \kappa \left(p - \frac{B^{t x} B_{t x}}{8 \pi} \right),  \label{gthetatheta}\\
 {G^{\phi}}_{\phi} & = & {G^{\theta}}_{\theta},
 \end{eqnarray}
where an overdot denotes differentiation with respect to the comoving time $t$, $\kappa= 8 \pi G /c^{4}$, and the coefficients $a_{0}$, $a_{1}$ and $a_{2}$ are given by:
\begin{eqnarray}
a_{0} & =& \frac{256 \; G Q^{2} c^{12} x^{4} a^{4}}{\left(G^{2}M^{2} -G Q^{2} + 4 G c^{2}M x a  + 4 c^{4} x^{2} a^{2}\right)^{4}},\\
a_{1} & = & \frac{- 5 G^{2} M^{2} + 5 G Q^{2} - 8G M c^{2} x a + 4 c^{4} x^{2} a^{2}}{-G^{2} M^{2} + G Q^{2} + 4 c^{4} x^{2} a^{2}},\\
a_{2} & = & \frac{G^{2}M^{2} - G Q^{2} + 4 G M c^{2} x a + 4 c^{4} x^{2} a^{2}}{-G^{2} M^{2} + G Q^{2} + 4 c^{4} x^{2} a^{2}}.  
\end{eqnarray}

We determine the explicit form of the tensor $B^{\mu \nu}$ subtracting Eq. \eqref{grr} from Eq. \eqref{gthetatheta}. After some algebraic manipulation, the non-zero components of $B^{\mu \nu}$ are:
\begin{equation}\label{bcomp}
B^{tx} = \frac{Q}{x^{2} a^{3} \left[1- \frac{G \left(G M^{2} - Q^{2}\right)}{4 c^{4} a^{2} x^{2}}\right] \left[\left(1 +\frac{G M}{2 c^{2} x a}\right)^{2} - \frac{G Q^{2}}{4 c^{4} a^{2} x^{2}}\right]^{2} },
\end{equation}
and $B^{xt} = -  B^{tx}$.

Next, we verify that Eq. \eqref{tb} is satisfied. Since $B^{\mu \nu}$ is an anti-symmetric tensor:
\begin{equation}
\nabla_{\mu} B^{\mu \nu} = \frac{1}{\sqrt{\lvert g \rvert}} \partial_{\mu} \left( \sqrt{\lvert g \rvert} B^{\mu \nu}\right) . 
\end{equation}
Furthermore, given that the only non-null components of $B^{\mu \nu}$ are $B^{t x}$ and $B^{xt}$, two of the four equations of \eqref{tb} are trivially satisfied. The other two remaining terms read:
\begin{eqnarray}
\frac{1}{\sqrt{\mid g \mid}} \partial_{x} \left( \sqrt{\mid g \mid} B^{x t}\right) & = &   \frac{1}{\sqrt{\mid g \mid}}\partial_{x} \left( Q \sin{\theta}\right) = 0,\\
\frac{1}{\sqrt{\mid g \mid}}\partial_{t} \left( \sqrt{\mid g \mid } B^{t x}\right) & = &   \frac{1}{\sqrt{\mid g \mid}}\partial_{t} \left( - Q \sin{\theta}\right) = 0.\\
\end{eqnarray}
Thus, Eq. \eqref{tb} holds.  On the other hand, it can be easily checked that the tensor $B^{\mu \nu}$ with components given by expression \eqref{bcomp} also satisfies Eq. \eqref{tb1}.

All these lengthy calculations were necessary to prove that there is an exact solution of STVG field equations that corresponds to an inhomogeneity in an expanding universe. Our next goal is to assess the nature of this spacetime; more specifically, we first analyze whether the metric becomes singular for a certain range of coordinates and values of the parameters.  Second, we compute the location of the apparent horizons and determine if they correspond to event or cosmological horizons. These features are essential to obtain a precise characterization of the spacetime and evaluate if cosmological black hole solutions are possible within the theory.


\section{\label{sec4}Properties of inhomogeneous expanding spacetimes in STVG}

It is convenient to express the line element \eqref{metric} in terms of the parameter $\alpha$:
\begin{equation}
ds^{2} = - c^{2}  \frac{{f(t,x)}^{2}}{{g(t,x)}^{2}} dt^{2} +{a(t)}^{2} {g(t,x)}^{2} \left(dx^{2} + x^{2} {d\Omega}^{2}\right),\label{metric2}
 \end{equation}
where
\begin{eqnarray}
f(t,x) & = &  \left[1- \frac{{G_{\rm N}}^{2} \left(1 + \alpha \right) M^{2}}{4 c^{4} {a(t)}^{2} x^{2}}\right],\\
g(t,x) & = &  \left[1 +\frac{G_{\rm N} \left(1+\alpha\right) M}{c^{2} x a(t)} + \frac{{G_{\rm N}}^{2} \left(1+\alpha\right) M^{2}}{4 c^{4} {a(t)}^{2} x^{2}}\right].
\end{eqnarray}

The limits of this metric are the expected: if $a \equiv 1$, \eqref{metric2} reduces to the line element of a Schwarzschild-STVG black hole written in isotropic coordinates, while in the limit $M \rightarrow 0$ Eq. \eqref{metric2} tends to the metric of a spatially flat FLRW model. For $\alpha \rightarrow 0$, the McVittie metric in GR is recovered.

\subsection{Singularities}\label{singularities}

Singularities are a pathological feature of some solutions of the fundamental equations of a theory \cite{rom13}. In GR and STVG, we can identify singular spacetime models if some physical quantity, for instance density or pressure of the fluid, or some curvature invariant is badly behaved. Thus, we begin computing the Ricci scalar for metric \eqref{metric2}:
\begin{eqnarray}
R  & = & R_{ab}R^{ab}  =  \frac{6}{f(t,x)}\left[\frac{\ddot{a}}{a} g(t,x) + {H(t)}^{2} \gamma\right],\label{ricci-scalar}\\
\gamma & = &   1 -  \frac{G_{\rm N} \left(1+\alpha\right) M}{c^{2} a(t) x} - \frac{3 {G_{\rm N}}^{2} \left(1 + \alpha\right) M^{2}}{4 c^{4} {a(t)}^{2} x^{2}}.
\end{eqnarray}
Inspection of the latter equation reveals that the Ricci scalar diverges if $f(t,x) = 0$, that is:
\begin{equation}\label{sing}
a(t) x = \frac{G_{\rm N} \left(1+\alpha\right)^{1/2} M}{2 c^{2}}.
\end{equation} 
According to the classification of spacetime singularities introduced by Ellis and Schmidt \cite{ell+77}, the metric possesses a scalar curvature singularity for those values of the coordinate $x$ that satisfy Eq. \eqref{sing}. In the limit $\alpha \rightarrow 0$, the singular points in McVittie metric are obtained.

The singularities of the metric corresponds to the hypersurface $\sigma(t,r) = 0$, where
\begin{equation}\label{singularity}
\sigma(t,r) =   a(t) x - \frac{G_{\rm N} \left(1+\alpha\right)^{1/2} M}{2 c^{2}}.
\end{equation}  
The normal vector $n_{a} =\nabla_{a} \sigma$ to the hypersurface and its corresponding  norm are \cite{poi04} :
\begin{equation}
n^{a}n_{a} = - \frac{1}{c^{2}} \frac{{g(t,x)}^2}{{f(t,x)}^{2}} {\dot{a}}^{2} x^{2} +\frac{1}{{g(t,x)}^{2}}. 
\end{equation}
Since in the limit $x \rightarrow G_{\rm N} \left(1+\alpha \right)^{1/2} M/ 2 c^{2} a(t)$, the norm of the normal vector tends to $- \infty$, the surface is spacelike, and consequently the singularity is spacelike.

We write the Ricci scalar in terms of the energy density of the fluid and its pressure: we take the trace of Eq. \eqref{fe1}, and using that $T^{\phi} = {T^{\mu}_{\mu}}^{\phi} = 0$, we get:
\begin{equation}\label{rel1}
  R = \frac{8 \pi G}{c^{4}}\left(\rho c^{2} - 3p\right).
\end{equation}
On the other hand, we obtain an additional relation between $\rho$ and $p$ by subtracting Eq. \eqref{gtt} from Eq. \eqref{grr}. The result is:
\begin{equation}\label{rel2}
 2 \dot{H(t)} \frac{g(t,x)}{f(t,x)} = -   \frac{8 \pi G}{c^{4}}\left(\rho c^{2} + p\right),
\end{equation}
Equations \eqref{rel1} and \eqref{rel2} form a system of two equations with two unknowns, $\rho$ and $p$. The solution is:
\begin{eqnarray}
\rho c^{2} & = & \frac{3 c^{4}}{8 \pi G_{\rm N} \left(1+\alpha\right)}{H(t)}^{2},\label{density} \\
p & = & - \frac{c^{4}}{8 \pi G_{\rm N} \left(1+\alpha\right)}\left[2 \; \dot{H(t)} \frac{g(t,x)}{f(t,x)} + 3 \; {H(t)}^{2}\right],\label{pressure}
\end{eqnarray}
In the limit $\alpha \rightarrow 0$, the corresponding expressions for the energy density and pressure in McVittie spacetime in GR are recovered \cite{far15}. Notice that $\rho$ is homogeneous on hypersurfaces of $t$ constant as opposed to the pressure. From expression \eqref{pressure}, we see that $p$ diverges in the same way as the Ricci scalar. Both the energy density and the pressure have the same qualitative features as in McVittie spacetime in GR \cite{kal+10}.



\subsection{Apparent horizons}

We characterize stationary black holes by the presence of event horizons. In dynamical spacetimes, however, to compute the location of the event horizon is an impossible task since we would need to know the entire spacetime manifold to future infinity. Instead,  we can resort to the concept of apparent horizon. This is defined as the boundary where the convergence properties of null geodesics congruences change. The apparent horizons are located where:
\begin{equation}
\theta_{n}  = 0,
\end{equation}
and
\begin{equation}
 \theta_{l}  > 0. 
\end{equation}
Here, $\theta_{n}$ and $\theta_{l}$ are the expansion of the future-directed ingoing and outgoing null geodesics congruences, respectively \cite{far15}. Apparent horizons are defined quasi-locally and do not refer to the global causal structure of spacetime \cite{far15}. In spherical symmetry,  the future-directed ingoing and outgoing null geodesics are radial and their tangent fields are denoted $n^{a}$ and $l^{a}$, respectively. If the null vectors $n^{a}$ and $l^{a}$ are not affinely-parametrized, their corresponding expansions are calculated as follows: 
\begin{equation}
 \theta_{n} = h^{a b} \nabla_{a}n_{b} = \left[g^{ab} + \frac{l^{a} n^{b} + n^{a} l^{b}}{\left(- n^{c} l^{d} g_{c d}\right)}\right] \nabla_{a}n_{b}, 
\end{equation}
where in the later equation $n_{b}$ should be substituted by $l_{b}$ in order to calculate $\theta_{l}$.  The tensor $h^{ab}$ acts as a projector onto the two-dimensional surface to which $n^{a}$ and $l^{a}$ are normal.

The tangent fields to the ingoing and outgoing radial null geodesics of metric \eqref{metric2} are:
\begin{eqnarray}
  n^{\mu} & = & \left(\frac{g(t,x)}{c \; f(t,x)}, \frac{-1}{a(t) g(t,x)}, 0, 0\right),\\
  l^{\mu} & = & \left(\frac{g(t,x)}{c \; f(t,x)}, \frac{1}{a(t) g(t,x)}, 0, 0\right).
  \end{eqnarray}
  These tangents fields are such that $n^{\mu}n_{\mu} = l^{\mu} l_{\mu} = 0$, and $g_{\mu \nu} n^{\mu}l^{\nu} = -2$. Given the vectors $n^{a}$ and $l^{b}$, after some algebraic manipulations, the expansions for $\theta_{n}$ and $\theta_{l}$ take the form:
  \begin{eqnarray}
  \theta_{n} & = & \frac{2 \gamma}{a(t) x {g(t,x)}^{2}}\left[x \dot{a}(t) \frac{g(t,x)}{c f(t,x)} - \frac{1}{g(t,x)}\right],\\
  \theta_{l} & = & \frac{2 \gamma}{a(t) x {g(t,x)}^{2}}\left[x \dot{a}(t) \frac{g(t,x)}{c f(t,x)} + \frac{1}{g(t,x)}\right],\\
    \gamma & = & 1+ \frac{G_{\rm N} M \left(1+\alpha\right)}{2 c^{2} x a(t)}.
  \end{eqnarray}
  
 The condition $\theta_{n} = 0$ implies $x \dot{a}(t) {g(t,x)}^{2} = c f(t,x)$, which in terms of the areal radius: 
 \begin{equation}\label{areal-radius}
R(t,x) = a(t) x  \left[1 +\frac{G_{\rm N} \left(1+\alpha\right) M}{c^{2} x a(t)} + \frac{{G_{\rm N}}^{2} \left(1+\alpha\right) M^{2}}{4 c^{4} {a(t)}^{2} x^{2}}\right],
 \end{equation} 
  can be written as:
  \begin{equation}\label{apparent-hor}
  \frac{{H(t)}^{2}}{c^{2}} R^{4} - R^{2} + 2 \frac{G_{\rm N} \left(1+\alpha\right) M}{c^{2}}R - \frac{{G_{\rm N}}^{2} \left(1+\alpha\right) \alpha M^{2}}{c^{4}} = 0,
  \end{equation}
and $H(t) = \dot{a(t)}/a(t)$ is the Hubble function. We can gain some insight on the nature of the apparent horizons of metric \eqref{metric2} by analyzing the limits of Eq. \eqref{apparent-hor}. For large values of the areal radius, $R \rightarrow c/H $; this is the value of the cosmological apparent horizon in the FLRW model. In the case $H \rightarrow 0$, Eq. \eqref{apparent-hor} reduces to:
\begin{equation}
  R^{2} - 2 \frac{G_{\rm N} \left(1+\alpha\right) M}{c^{2}}R + \frac{{G_{\rm N}}^{2} \left(1+\alpha\right) \alpha M^{2}}{c^{4}} = 0.
\end{equation}
The two solutions of the quadratic equation are:
  \begin{equation}
  R_{\pm} = \frac{G_{\rm N} M}{c^{2}} \left[\left(1 + \alpha\right) \pm \left(1 + \alpha\right)^{1/2} \right]  .
 \end{equation}
  These are the outer $(+)$ and  inner $(-)$ event horizons in the Schwarzschild STVG black hole \cite{mof15a}. Finally, if we take $\alpha \rightarrow 0$, Eq. \eqref{apparent-hor} reduces to a cubic equation that locates the apparent horizons in McVittie metric in GR (see for instance Equation (4.25) in \cite{far15}). Thus, in the appropriate limits, the apparent horizons become a cosmological or a black hole event horizon, a strong hint that this metric may represent a cosmological black hole.
  
\begin{figure}
\center
   \resizebox{\hsize}{!}{\includegraphics{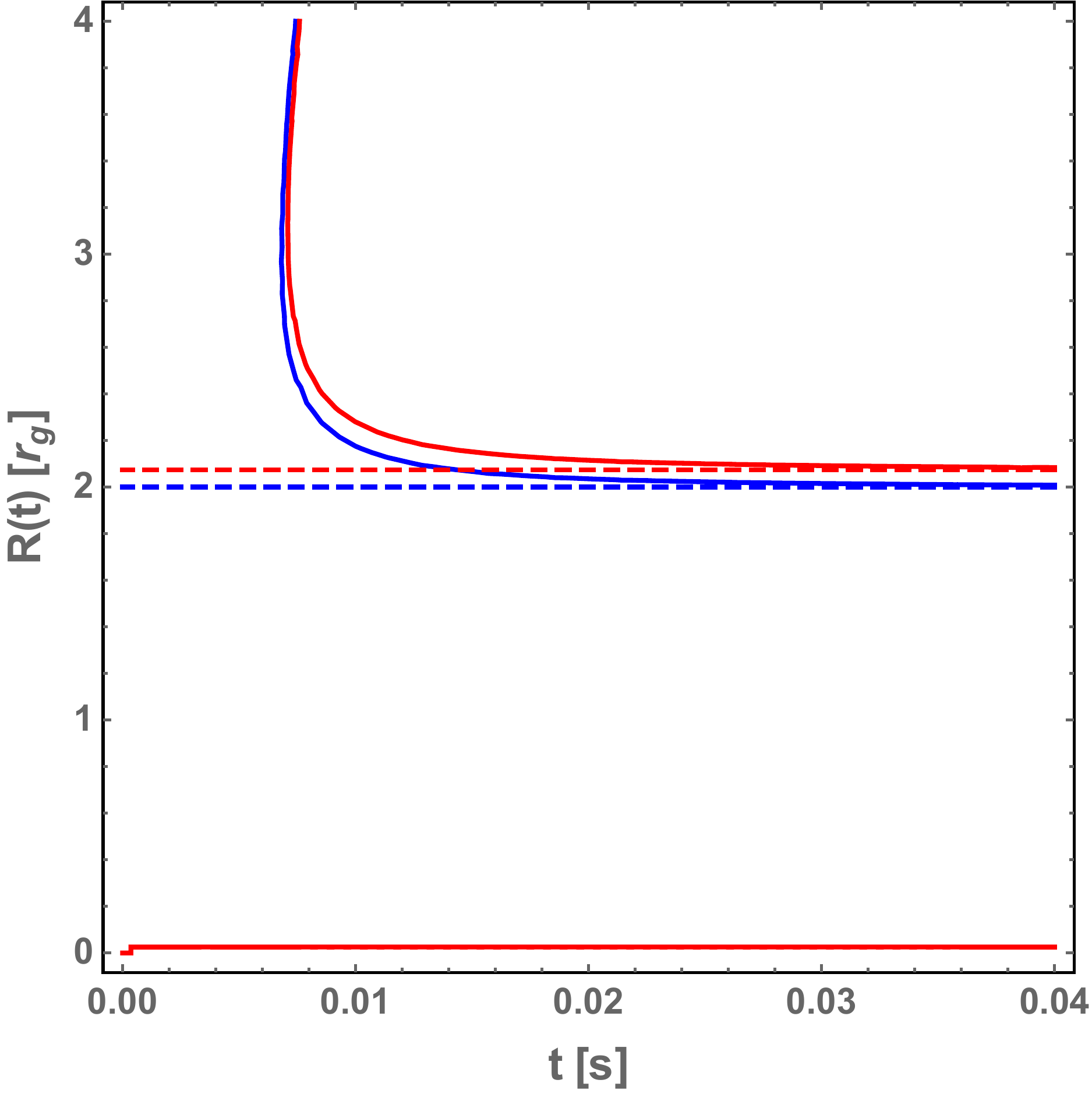}}
\caption{Plot of the  areal radius of the apparent horizons as a function of time for a stellar mass source in a dust-dominated background. The blue line corresponds to $\alpha = 0$, and the red line to $\alpha = 5 \times 10^{-2}$. For each case, the dashed line indicates the location of the singularity.}
\label{fig1}
 \end{figure}

\begin{figure}
\center
   \resizebox{\hsize}{!}{\includegraphics{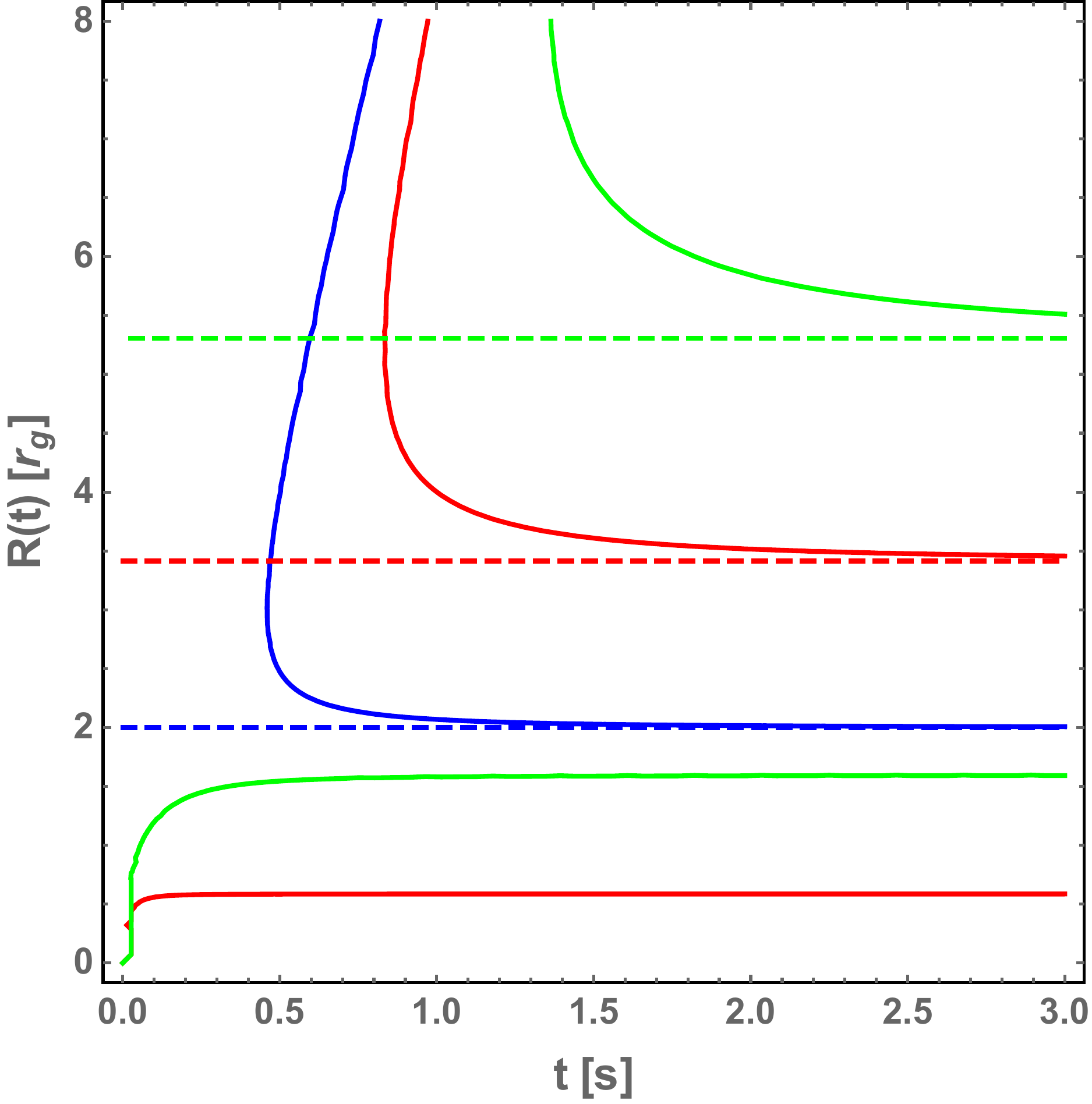}}
\caption{Plot of the  areal radius of the apparent horizons as a function of time for a supermassive black hole  in a dust-dominated background. The blue line corresponds to $\alpha = 0$, the red line to $\alpha = 1$, and the green line $\alpha = 2.45$. For each case, the dashed line indicates the location of the singularity.}
\label{fig2}
 \end{figure}

\begin{figure}
\center
   \resizebox{\hsize}{!}{\includegraphics{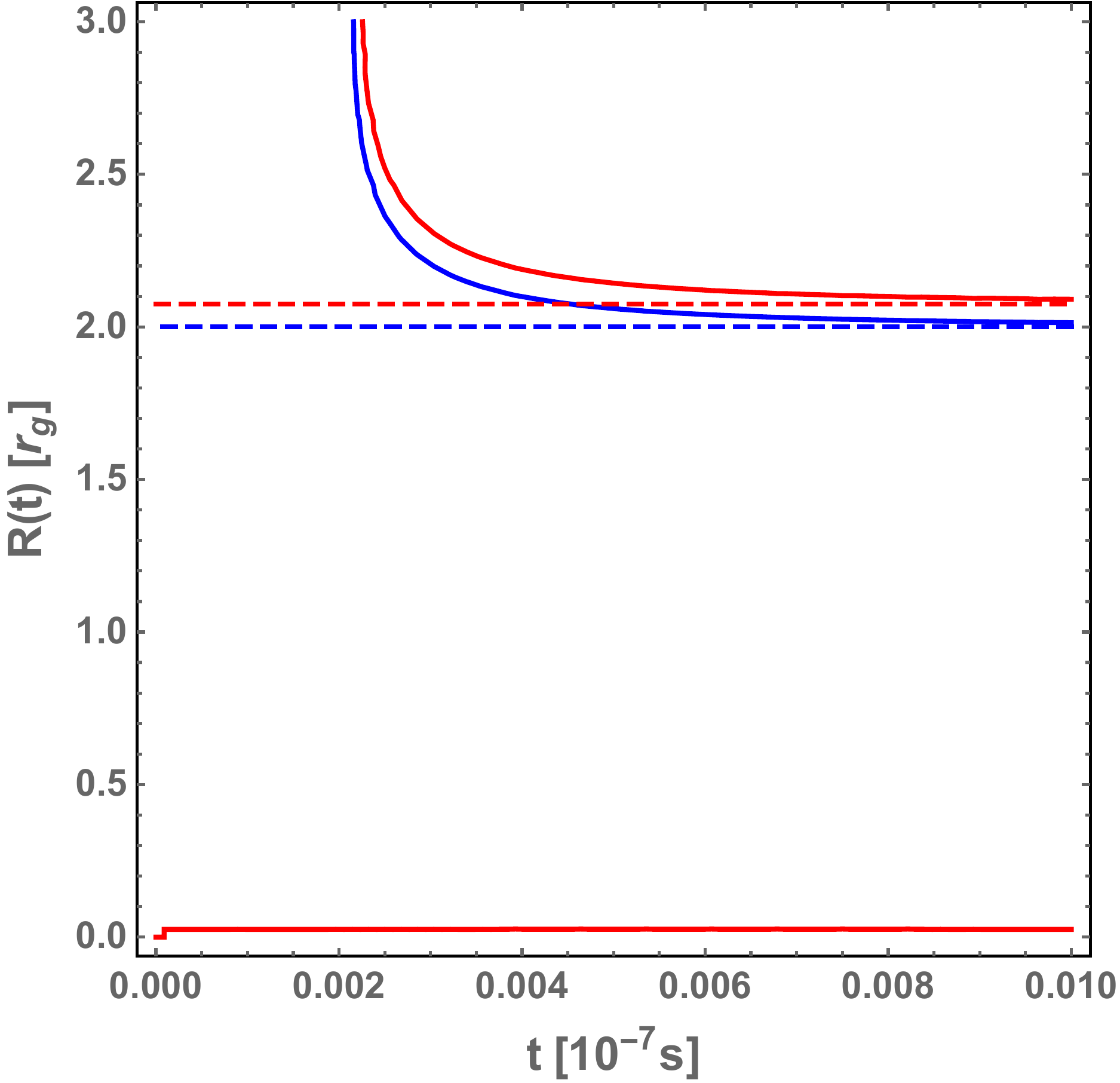}}
\caption{Plot of the  areal radius of the apparent horizons as a function of time for a stellar mass source in the $\Lambda$-CDM model. The blue line corresponds to $\alpha = 0$, and the red line to $\alpha = 5 \times 10^{-2}$. For each case, the dashed line indicates the location of the singularity.}
\label{fig3}
 \end{figure}

\begin{figure}
\center
   \resizebox{\hsize}{!}{\includegraphics{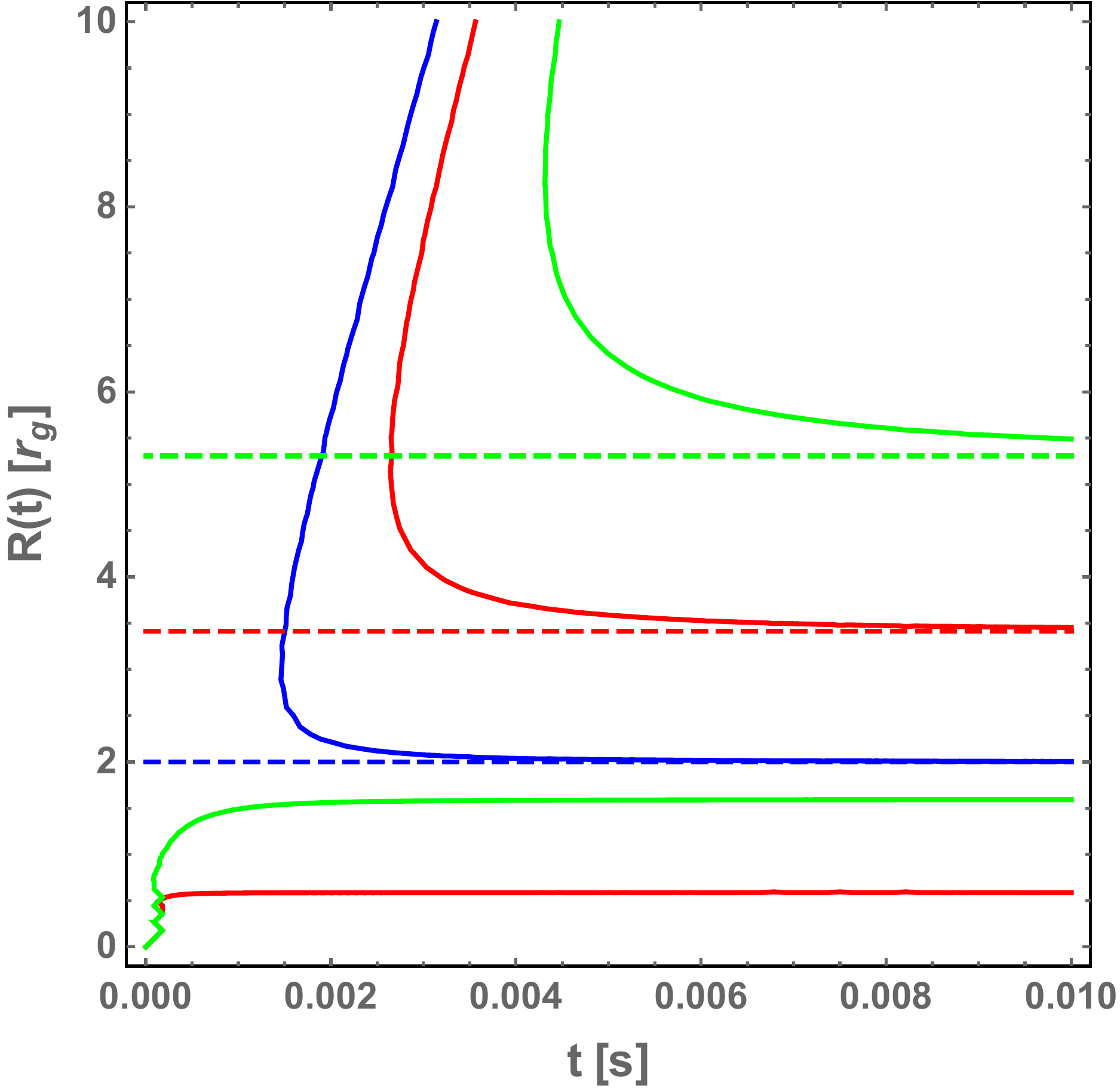}}
\caption{Plot of the  areal radius of the apparent horizons as a function of time for a supermassive black hole the $\Lambda$-CDM model. The blue line corresponds to $\alpha = 0$, the red line to $\alpha = 1$, and the green line $\alpha = 2.45$. For each case, the dashed line indicates the location of the singularity.}
\label{fig4}
 \end{figure}

Equation \eqref{apparent-hor}, nonetheless, has four roots. Using Descartes' rule of sign, we determine that three of them are positive and one is negative. We discard the latter because it has no physical meaning. Let us denote the three positive roots $R_{*}$, $R_{-}$, and $R_{+}$, where $R_{*} < R_{-} < R_{+}$. We show plots of the roots as a function of the cosmic time from Figures \ref{fig1} to \ref{fig4}.  For Figures \ref{fig1} and \ref{fig2}, the Hubble function is that of a cosmological dust dominated background model:
\begin{equation}
 H(t) = \frac{2}{3} \frac{1}{t}. 
\end{equation}
In Figures \ref{fig3} and \ref{fig4}, we adopt the scale factor of the $\Lambda$ Cold Dark Matter model ($\Lambda$CDM):
\begin{equation}\label{eq:sflambda}
 a(t) =  \left[\frac{\left(1 - \Omega_{\Lambda, 0}\right)}{\Omega_{\Lambda, 0}} \left(\sinh{\left(\frac{3}{2} H_{0} \sqrt{\Omega_{\Lambda, 0}} \; t\right)}\right)^{2}\right]^{1/3}.
\end{equation}
 Here, $H_{0} = 2.27 \times 10^{-18} {\mathrm s^{-1}}\approx 70$ km/s Mpc, and $\Omega_{\Lambda, 0} = 0.7$ for the Hubble factor and the cosmological constant density parameter, respectively.

The value of the parameter $\alpha$ depends on the mass of the gravitational central source. For stellar mass sources,  Lopez Armengol and Romero \cite{lop+17a} found that $\alpha < 0.1$. In the case of supermassive black holes ($10^{7} M_{\odot} \le M \le 10^{9} M_{\odot}$) the range of values are $0.03 < \alpha < 2.47$ (see for instance \cite{mof+08}, \cite{bro+06b}, and \cite{per+17}). Figures \ref{fig1} and \ref{fig3} correspond to a stellar mass source (we choose $\alpha = 5 \times 10^{-2}$) while for Figures \ref{fig2} and \ref{fig4} the source is a supermassive black hole (we select $\alpha = 1$, and $\alpha = 2.45$). In the four plots, we include the apparent horizons in McVittie spacetime in GR ($\alpha = 0$) for comparison.

There are common features to all these plots:
\begin{itemize}
\item  We distinguish three apparent horizons. Two of them, $R_{-}$ and $R_{+}$, lay in the causal future of the curvature singularity (see Eq. \eqref{singularity}). The innermost apparent horizon $R_{*}$ is  bounded by the singularity and disconnected from the exterior geometry. In what follows, we restrict our analysis to the spacetime region that corresponds to the causal future of the curvature singularity.
\item The curvature singularity (dashed line in the Figures) is present since $t = 0$ and its location in terms of the areal radius does not change with  cosmic time. Since the surface given by Eq. \eqref{singularity} at $t = 0$ is in the causal past of all the spacetime events of the region of interest, we regard it as a cosmological ``Big-Bang'' singularity\footnote{Kaloper and collaborators \cite{kal+10} give the same interpretation for the curvature singularity in McVittie spacetime in GR.}.
\item At early values of the cosmic time, we only have a cosmological singularity. Later on, the apparent horizons $R_{-}$ and $R_{+}$ appear together at a specific value of the cosmic time. The horizon $R_{+}$ becomes larger for growing $t$, reaching the value of the cosmological apparent horizon in the FLRW model.  Conversely, $R_{-}$ gets smaller for increasing values of $t$, and in the limit $t \rightarrow \infty$, it gets closer and closer to the singularity. 
\item For larger values of the parameter $\alpha$, the appearance of the apparent horizons occurs at later times. Furthermore, the value of the areal radius of the surface singularity and the horizons is higher for increasing values of $\alpha$.
\end{itemize}

Since $R_{+}$ expands forever and  it tends to the cosmological apparent horizon in the FLRW model, we interpret the surface $R_{+}$, $t$ finite as a cosmological apparent horizon of the spacetime metric \eqref{metric}.
  
  The nature of the apparent horizon $R_{-}$ requires some further analysis. We are particularly interested in the surface $R = R_{-}$, $t = \infty$.  In the next subsection, we show that independently of the asymptotic form of the Hubble function as long as the null energy condition is satisfied, ingoing null radial geodesics reach the null surface $R = R_{-}$, $t = \infty$ in a finite lapse of the affine parameter. In other words, the spacetime metric \eqref{metric} is incomplete to null future-oriented ingoing geodesics\footnote{In other to obtain this result, we follow the methods developed by Kaloper at al. \cite{kal+10}.}.
  
 \subsubsection{The surface $R = R_{-}$, $t = \infty$} 
 
 Consider ingoing null radial geodesics from an initial distance $R_{\rm i} > R_{-}$ and geodesic initial velocity at that point $R'_{\rm i} < 0$. We aim to show that these geodesics arrive in a finite lapse of affine parameter $\sigma$ to the surface $R = R_{-}$, $t = \infty$. More precisely:
 \begin{equation}
R' = \frac{dR}{d\sigma} \Rightarrow d\sigma = \frac{dR}{R'} \Rightarrow   \Delta \sigma = \int_{\sigma_{R_{-}}}^{\sigma} d\sigma = \int_{R_{-}}^{R} \frac{dR}{R},
 \end{equation} 
 being $\Delta \sigma$ a finite quantity.
 
   Before starting, it is convenient to express the line element  \eqref{metric2} in terms of the areal radius (see Eq. \eqref{areal-radius}). When doing the coordinate transformation, the algebraic manipulations are considerably simplified if you employ the relation:
 \begin{equation}\label{rel}
 \frac{{f(t,x)}^{2}}{{g(t,x)}^{2}} = 1 - \frac{4 r_{0}}{R} + \frac{4 {r_{1}}^{2}}{R^{2}},
  \end{equation} 
where we introduce $r_{0}$ and $r_{1}$ to simplify the notation:
\begin{eqnarray}
  r_{0} & = & \frac{G_{\rm N} \left(1+\alpha\right) M}{2 c^{2}},\\
  r_{1} & = & \frac{G_{\rm N} \sqrt{\left(1+\alpha\right) \alpha} M}{2 c^{2}}.
\end{eqnarray}
After the coordinate transformation, the line element takes the form:
\begin{eqnarray}
ds^{2} & = & - c^{2} \left(1-  \frac{4 r_{0}}{R} + \frac{4 {r_{1}}^{2}}{R^{2}} - \frac{{H(t)}^{2} R^{2}}{c^{2}}\right) dt^{2} \\ 
& - & \frac{2 R H(t)}{\sqrt{1-  \frac{4 r_{0}}{R} + \frac{4 {r_{1}}^{2}}{R^{2}}}} dR dt +  \frac{{dR}^{2}}{1-  \frac{4 r_{0}}{R} + \frac{4 {r_{1}}^{2}}{R^{2}}} + R^{2} {d\Omega}^{2}.\nonumber
\end{eqnarray}

For radial null geodesics, $ds^{2} = 0$ and we derive the equation:
\begin{equation}\label{geo1}
\frac{dt}{dR}\Bigg|_{\pm} = \frac{\pm}{c}\frac{1}{\sqrt{1 - \frac{4 r_{0}}{R} + \frac{4 {r_{1}}^{2}}{R^{2}}} \pm \frac{H(t) R}{c}}  \frac{1}{\sqrt{1 - \frac{4 r_{0}}{R} + \frac{4 {r_{1}}^{2}}{R^{2}}}}.
\end{equation}
Here, the $+$ ($-$) sign corresponds to outgoing (ingoing) radial null geodesics, respectively. 

We focus on ingoing radial geodesics and rewrite Eq. \eqref{geo1} as:
\begin{eqnarray}
 \frac{dt}{dR} & = & - \frac{1}{c}\frac{\sqrt{1 - \frac{4 r_{0}}{R} + \frac{4 {r_{1}}^{2}}{R^{2}}} + \frac{H(t) R}{c}}{\sqrt{1-  \frac{4 r_{0}}{R} + \frac{4 {r_{1}}^{2}}{R^{2}}} \left(1-  \frac{4 r_{0}}{R} + \frac{4 {r_{1}}^{2}}{R^{2}} - \frac{{H(t)}^{2} R^{2}}{c^{2}}\right)}\nonumber \\
 & = & - \frac{R^{2}}{{H(t)}^{2}}\frac{\left(1 + \frac{H(t) R}{c \sqrt{1 - \frac{4 r_{0}}{R} + \frac{4 {r_{1}}^{2}}{R^{2}}}}\right)}{\left(R - R_{*}\right)  \left(R_{+} - R\right)\left(R +R_{*} + R_{-} + R_{+}\right)}\nonumber \\
 & & \times \frac{1}{\left(R - R_{-}\right)}.
\end{eqnarray}

In the limit $R \rightarrow R_{-}$,
\begin{eqnarray}
dt & \rightarrow & -   \frac{{R_{-}}^{2}}{{\tilde{H_{0}}}^{2}} \frac{1}{\left(R_{-} - R_{*}\right)  \left(R_{+} - R_{-}\right)\left(2 R_{-}+(R_{*} + R_{+}\right)} \nonumber \\
& & \times \frac{dR}{\left(R - R_{-}\right)}.
\end{eqnarray}
$\tilde{H_{0}}$ is the value of the Hubble function in the limit $t \rightarrow \infty$. To leading order, integration of the latter yields:
\begin{equation}\label{result1}
e^{\left(\tilde{H_{0}}t\right)} \rightarrow \left(\frac{1}{R - R_{-}}\right)^{\gamma}  + ...,
\end{equation}
where $\gamma = {R_{-}}^{2}/\left[\left(R_{-} - R_{*}\right) \left(R_{+} - R_{-}\right)\left(2 R_{-}+(R_{*} + R_{+}\right)\right]$, being $\gamma > 0$.  Later on, we will use the result given by \eqref{result1}.

Next, we compute an additional radial null geodesic equation. After some algebra, we obtain\footnote{The equation for $R''$ can be derived from Lagrange equations, defining the Lagrangian $\mathcal{L} = F_{+} F_{-}$, where:
\begin{eqnarray}
F_{+} & = & c \left(\beta^{1/2} - \frac{H(t)R}{c}\right)t'+\frac{R''}{\beta^{1/2}},\\
F_{-} & = &   c \left(-\beta^{1/2} - \frac{H(t)R}{c}\right)t'+\frac{R''}{\beta^{1/2}},\\
\beta & = & 1 - 4 \frac{r_{0}}{R} + \frac{4 {r_{1}}^{2}}{R^{2}}.
\end{eqnarray}
After computing the Lagrange equations, it should be set $F_{+} = 0$ to derive the equations that corresponds to radial ingoing null geodesics \cite{kal+10}.}:
\begin{equation}\label{acceleration}
 R'' = \frac{R \dot{H(t)} {R'}^{2}}{c^{2} \sqrt{1 - \frac{4 r_{0}}{R} + \frac{4 {r_{1}}^{2}}{R^{2}}} \left(\sqrt{1 - \frac{4 r_{0}}{R} + \frac{4 {r_{1}}^{2}}{R^{2}}} - \frac{H(t) R}{c}\right)^{2}}. 
\end{equation}
The primes denote the derivative with respect to some affine parameter $\sigma$.

Notice that $\dot{H(t)} < 0$ provided the null energy condition holds. This can be proved by adding Eqs. \eqref{density} and \eqref{pressure}:
\begin{equation}
 \rho c^{2} + p = \frac{- c^{4}}{4 \pi G_{\rm N}} \left(1+\alpha\right) \dot{H(t)} \frac{g(t,x)}{f(t,x)}. 
\end{equation}
We see that if the null energy condition is satisfied, $\rho c^{2} +p > 0$, then $\dot{H(t)} < 0$. The following step is to look for an approximated formula for $\dot{H(t)}$: substituting an expression for the energy density of the universe:
\begin{equation}
\rho = \Lambda + \rho_{0} \left(\frac{a_{0}}{a(t)}\right)^{s},  
\end{equation}
where $s = 3 \left(1 + w\right)$ and $w = p/\rho$, into Eq. \eqref{density} and also taking into account the definition of the Hubble function, $\dot{H(t)} = \dot{a(t)}/a(t)$, in the limit $t \rightarrow \infty$, $H(t) \rightarrow \tilde{H_{0}} + \mathcal{O}\left(e^{\left(- s \tilde{H_{0}} t\right)}\right)$, and hence $\dot{H(t)} \propto e^{\left(- s \tilde{H_{0}} t\right)}$. Now, we make use of the approximation given in \eqref{result1}:
\begin{equation}
 \dot{H(t)} \propto (R - R_{*})^{\gamma s}. 
\end{equation}

Then, along null ingoing radial geodesics near the surface $R = R_{-}$, $t = \infty$:
\begin{equation}
 R'' \rightarrow - \tilde{\mathcal{C}} \left(R - R_{-}\right)^{\gamma s -2} {R'}^{2}. 
\end{equation}
Integration of the latter to leading order yields:
\begin{eqnarray}
R' & = & R'_{\rm i}  e^{\left(- \int_{R_{\rm i}}^{R} \tilde{\mathcal{C}} \left(R - R_{-}\right)^{\gamma s -2} \right)} dR \nonumber \\
& = &   \mathcal{G} e^{- \frac{\tilde{\mathcal{C}}}{\left(\alpha s -1\right)} \left(R - R_{-}\right)^{\gamma s -1}}.\label{vel}
\end{eqnarray}
We denote by $R'_{\rm i}$ the initial radial velocity of an ingoing geodesics that begins at the areal radius $R = R_{\rm i} > R_{-}$; we also consider that $R'_{\rm i} < 0$. Under this assumption, the constant $\mathcal{G}$ is finite and negative:
\begin{equation}
\mathcal{G} = R'_{\rm i} e^{ \frac{\tilde{\mathcal{C}}}{\left(\gamma s -1\right)}} \left(R_{\rm i} - R_{-}\right)^{\gamma s -1}.
\end{equation}

Using Eq. \eqref{vel}, we can straightforward estimate the quantity $\Delta \sigma$:
\begin{equation}
\Delta \sigma \; \mathcal{G} = \int_{R_{\rm i}}^{R_{-}} \; dR \;  e^{ \frac{\tilde{\mathcal{C}}}{\left(\gamma s -1\right)} \left(R - R_{-}\right)^{\gamma s -1}}.
  \end{equation}
  If $\alpha s -1 < 0$, then
  \begin{equation}
  \lim_{R \rightarrow R_{-}}   \frac{\tilde{\mathcal{C}}}{\left(\gamma s -1\right)} \left(R - R_{-}\right)^{\gamma s -1} \rightarrow - \infty,
  \end{equation}
and the integral is convergent.  In the case $\gamma s -1 \ge 0$:
\begin{equation}
  \lim_{R \rightarrow R_{-}}   \frac{\tilde{\mathcal{C}}}{\left(\gamma s -1\right)} \left(R - R_{-}\right)^{\gamma s -1} \rightarrow 0,
  \end{equation}
the integral is also convergent. Consequently, the integral always remains finite and so the quantity $\Delta \sigma$ \footnote{In the case the asymptotic value of $H(t)$ vanishes, the demonstration exists and is quite alike.}.  Hence, we have proved that ingoing radial null geodesics arrive at the surface $R = R_{-}$, $t = \infty$ in a finite lapse of affine parameter.

The event horizon is characterized as a one way membrane: once we have crossed this surface it is physically impossible to cross it back in the opposite sense. This is precisely the case for the surface  $R = R_{-}$, $t = \infty$ provided $\tilde{H_{0}} > 0$ when $t \rightarrow \infty$. 

Consider again a radial ingoing null geodesic with initial velocity $R' <0$. According to Eq. \eqref{acceleration}, and since $\dot{H(t)} < 0$, the acceleration $R''$ is negative. This geodesic can never turn back or decrease its speed. Once the geodesic arrives at $R = R_{-}$, $t = \infty$ in a finite lapse of affine parameter, it crosses this surface which is perfectly traversable. Recall that $R = R_{-}$ is a null branch of the apparent horizon, and thus constitutes a boundary where the convergence properties of null geodesics change. Right after crossing $R = R_{-}$, $t = \infty$, the convergence properties of the geodesic are modified  and it is unable to return back. Hence, the surface $R = R_{-}$, $t = \infty$ is an event horizon, and the spacetime metric \eqref{metric2} represents a cosmological black hole.

The nature of the surface $R = R_{-}$, $t = \infty$ when $\tilde{H_{0}} = 0$ is much more subtle. The equation for the location of the apparent horizons \eqref{apparent-hor} can be rewritten as:
\begin{equation}
f_{\mathrm ah} = 1 - 4 \frac{r_{0}}{R} + 4 \frac{{r_{1}^{2}}}{R^{2}} - \frac{{H(t)}^2 R^{2}}{c^{2}}.  
\end{equation}
In the limit $t \rightarrow \infty$, $H(t) \rightarrow 0$, and $f_{\rm ah}$ reduces to:
\begin{equation}
 1 - \frac{4 r_{0}}{R} + \frac{4 {r_{1}}^{2}}{R^{2}} = \frac{{f(t,x)}^{2}}{{g(t,x)}^{2}},
\end{equation}
where we employ the equality given by \eqref{rel}. The apparent horizons are located where $f_{\rm ah}  =0$, or equivalently where $f(t,x) = 0$. As shown in Section \ref{singularities}, $f(t,x) = 0$ identifies a singular surface of the spacetime metric. Consequently, the cosmological solution does not represent a black hole. Further investigation is needed to assess the strength of the singularity, and thus to get a better understanding of the global causal structure of the spacetime.


\section{Discussion}\label{sec5}

Until now, we have explored the properties of the solution for a limited range of values of the parameter $\alpha$. In what follows, we remove such restriction and allow $\alpha$ to freely move in the interval $0 < \alpha < \infty$.

As already mentioned in Section \ref{singularities}, the location of the singularity is independent of the cosmic time. We write Eq. \eqref{singularity} in terms of the areal radius:
\begin{equation}
R \sqrt{1+\alpha} - \left(1+\alpha\right) \left[1 + \sqrt{1+\alpha}\right] = 0,
\end{equation} 
or
\begin{equation}
R(\alpha)= \sqrt{1+\alpha} \left[1 + \sqrt{1+\alpha}\right].
\end{equation}
The function $R(\alpha)$ is strictly increasing: the higher the value of $\alpha$, the larger the areal radius of the singularity. Also notice that there is no value for $\alpha$ such that the singularity can be avoided, implying that there are no regular cosmological black hole solutions in the theory.

\begin{figure}
\center
   \resizebox{\hsize}{!}{\includegraphics{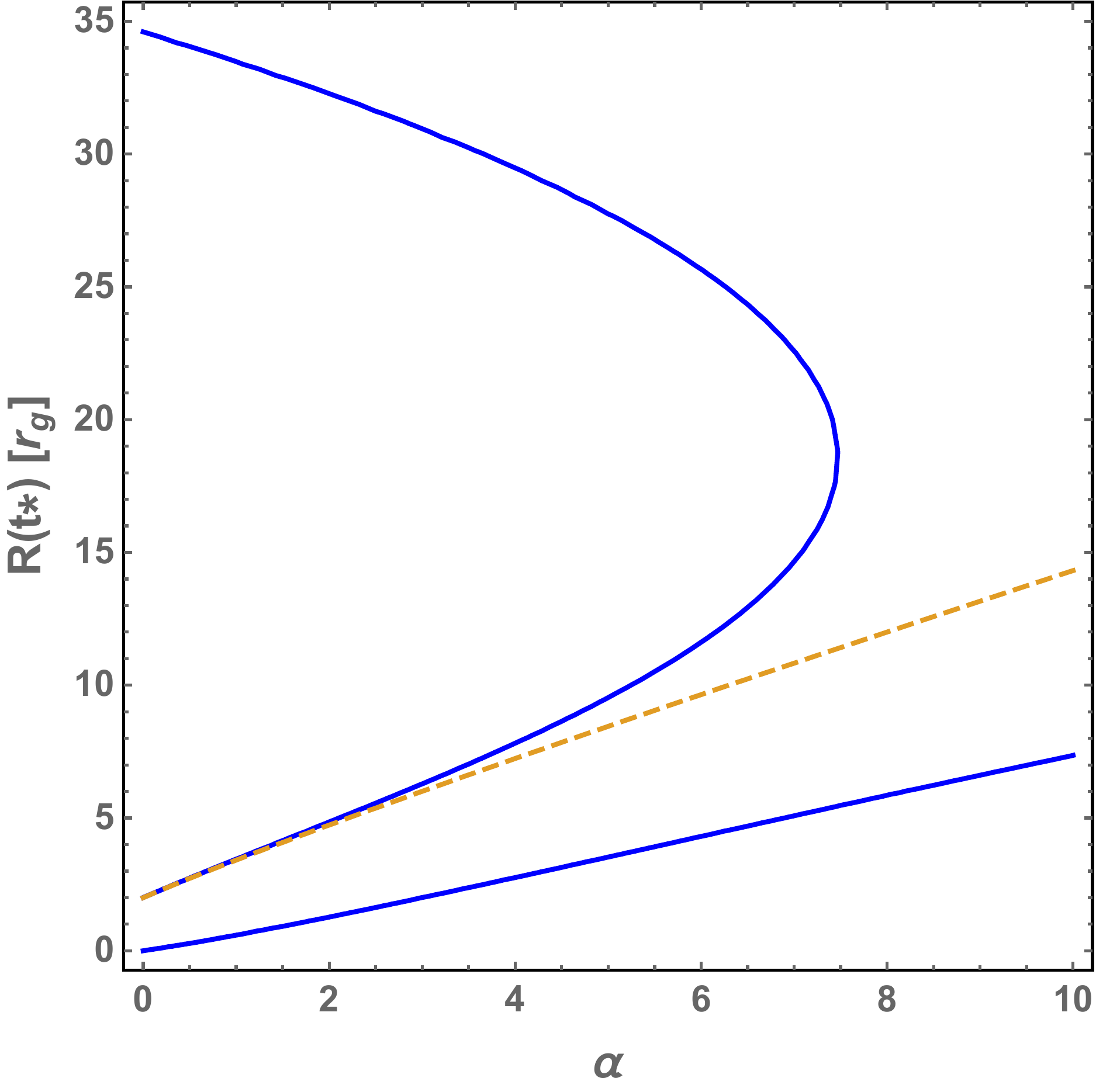}}
\caption{Plot of the  areal radius of the apparent horizons as a function of the parameter $\alpha$ for a fixed value of the cosmic time. The central source is a  supermassive black hole, and the Hubble factor corresponds to the $\Lambda$-CDM model. The dashed line indicates the location of the singularity.}
\label{fig5}
 \end{figure}

The location of the apparent horizons as a function of the parameter $\alpha$ is depicted in Figure \ref{fig5}, for a fixed value of the cosmic time. As before, the dashed line marks the location of the singularity. In the interval $0 < \alpha < \tilde{\alpha}$, we identify three apparent horizons: an inner horizon $R_{*}$ that lies beyond the singularity (and hence is not part of the spacetime), a black hole apparent horizon $R_{-}$, and a cosmological apparent horizon $R_{+}$. As $\alpha$ gets closer to $\tilde{\alpha}$, $R_{-}$ increases while $R_{+}$ becomes smaller. For $\alpha = \tilde{\alpha}$ both horizons, $R_{-}$ and $R_{+}$, become one.

Higher values of $\alpha$ implies an augmented gravitational constant. In STVG the gravitational field is stronger that in GR; the central source drags the cosmological horizon while the black hole apparent horizon enlarges.

If $\alpha > \tilde{\alpha}$, the apparent horizons disappear and a naked singularity is left behind. Accepting the validity of the cosmic censorship conjecture \cite{pen69}, we see that restrictions can be imposed on the values of the parameter $\alpha$ such that solutions that contain naked singularities are not allowed in the theory. The constraint on $\alpha$ changes for different values of the cosmic time  (the coefficients of Eq. \eqref{apparent-hor} depend on the Hubble function $H(t)$). The latter implies that the permitted values of $\alpha$ do not only depend on the mass of the central source but also on the cosmic epoch of the universe.


\section{Conclusions}\label{sec6}

In this work we derive the first exact solution of STVG field equations that represents an inhomogeneity in an expanding universe. When the Hubble factor is positive at late cosmic times, we prove that the metric describes a black hole immersed in a cosmological background.

The spacetime presents a spacelike singular surface where the pressure of the cosmological fluid diverges, a feature that is common to McVittie metric in GR. We also show that there is no value of the parameter $\alpha$ of the theory such that the singularity can be avoided. This result implies that there are no regular cosmological black hole solutions in STVG.

The metric has two apparent horizons: an inner horizon and an outer horizon that correspond to an event and cosmological horizon for the black hole case. As the value of the parameter $\alpha$ increases, the size of the horizons enlarges as well as the areal radius that locates the singular surface.

We show that for both the $\Lambda$-CDM and the cosmological dust dominated background models, the apparent horizons begin to exist together and, as time goes by, their separation enlarges. The inner horizon approaches the singularity while the outer one tends to the cosmological horizon in the FLRW model.

For a given value of the cosmic time, there is a limited range of values of $\alpha$ such that the solution exhibits  an inner and an outer apparent horizon. Beyond this range, both horizons merge and finally disappear  leaving behind a naked singularity. If we assume the validity of the cosmic censorship conjecture, we see that only some values of $\alpha$ are allowed. Thus, the value of $\alpha$ is not only dependent on the mass of the central source but on the cosmic epoch. This result should be taken into account when modeling the evolution of the structure and the dynamics of astrophysical systems through cosmic time.

This work is a first step towards a better understanding of cosmological black holes in STVG. There are several issues that remain unexplored; for instance, the strength of the spacelike surface singularity, the nature of the cosmological solution when the Hubble factor is zero at late times, the dynamics of particles in this spacetime, just to mention some. The fact that STVG admits cosmological black hole solutions is yet another positive indicator that  the theory offers a suitable classical description of the various manifestations of gravity.

\begin{acknowledgments}
This work was supported by grants AYA2016-76012-C3-1-P (Ministerio de Educaci\'on, Cultura y Deporte, Espa\~{n}a) and PIP 0338 (CONICET, Argentina).     
  \end{acknowledgments}

\bibliographystyle{JHEP}
\bibliography{Perez-Romero}

\providecommand{\href}[2]{#2}\begingroup\raggedright\begin{thebibliography}{10}

\bibitem{lan+71}
L.~D. {Landau} and E.~M. {Lifshitz}, \emph{{The classical theory of fields}}.
  1971.

\bibitem{rom18}
G.~E. {Romero}, \emph{{Outline of a Theory of Scientific Aesthetics}},
  {\emph{Foundations of Science} {\bfseries 23} (2018) 795}
  [\href{https://arxiv.org/abs/1210.2427}{{\ttfamily 1210.2427}}].

\bibitem{ein15b}
A.~{Einstein}, \emph{{The Collected Papers of Albert Einstein, Volume 6: The
  Berlin Years: Writings, 1914-1917.}} 1996.

\bibitem{apr+12}
E.~{Aprile}, M.~{Alfonsi}, K.~{Arisaka}, F.~{Arneodo}, C.~{Balan}, L.~{Baudis}
  et~al., \emph{{Dark Matter Results from 225 Live Days of XENON100 Data}},
  \href{https://doi.org/10.1103/PhysRevLett.109.181301}{\emph{Physical Review
  Letters} {\bfseries 109} (2012) 181301}
  [\href{https://arxiv.org/abs/1207.5988}{{\ttfamily 1207.5988}}].

\bibitem{lux+13}
{LUX Collaboration}, D.~S. {Akerib}, H.~M. {Araujo}, X.~{Bai}, A.~J. {Bailey},
  J.~{Balajthy} et~al., \emph{{First results from the LUX dark matter
  experiment at the Sanford Underground Research Facility}}, {\emph{ArXiv
  e-prints} (2013) } [\href{https://arxiv.org/abs/1310.8214}{{\ttfamily
  1310.8214}}].

\bibitem{agn+14}
R.~{Agnese}, A.~J. {Anderson}, M.~{Asai}, D.~{Balakishiyeva}, R.~{Basu Thakur},
  D.~A. {Bauer} et~al., \emph{{Search for Low-Mass Weakly Interacting Massive
  Particles with SuperCDMS}},
  \href{https://doi.org/10.1103/PhysRevLett.112.241302}{\emph{Physical Review
  Letters} {\bfseries 112} (2014) 241302}
  [\href{https://arxiv.org/abs/1402.7137}{{\ttfamily 1402.7137}}].

\bibitem{mof06}
J.~W. {Moffat}, \emph{{Scalar tensor vector gravity theory}},
  \href{https://doi.org/10.1088/1475-7516/2006/03/004}{\emph{J. Cosmol.
  Astropart. Phys.} {\bfseries 3} (2006) 004}
  [\href{https://arxiv.org/abs/gr-qc/0506021}{{\ttfamily gr-qc/0506021}}].

\bibitem{bro+06a}
J.~R. {Brownstein} and J.~W. {Moffat}, \emph{{Galaxy Rotation Curves without
  Nonbaryonic Dark Matter}}, \href{https://doi.org/10.1086/498208}{\emph{The
  Astrophysical Journal} {\bfseries 636} (2006) 721}
  [\href{https://arxiv.org/abs/astro-ph/0506370}{{\ttfamily
  astro-ph/0506370}}].

\bibitem{mof+13}
J.~W. {Moffat} and S.~{Rahvar}, \emph{{The MOG weak field approximation and
  observational test of galaxy rotation curves}},
  \href{https://doi.org/10.1093/mnras/stt1670}{\emph{Mon. Not. R. Astron. Soc.}
  {\bfseries 436} (2013) 1439}
  [\href{https://arxiv.org/abs/1306.6383}{{\ttfamily 1306.6383}}].

\bibitem{mof+15}
J.~W. {Moffat} and V.~T. {Toth}, \emph{{Rotational velocity curves in the Milky
  Way as a test of modified gravity}},
  \href{https://doi.org/10.1103/PhysRevD.91.043004}{\emph{Physical Review D}
  {\bfseries 91} (2015) 043004}
  [\href{https://arxiv.org/abs/1411.6701}{{\ttfamily 1411.6701}}].

\bibitem{bro+06b}
J.~R. {Brownstein} and J.~W. {Moffat}, \emph{{Galaxy cluster masses without
  non-baryonic dark matter}},
  \href{https://doi.org/10.1111/j.1365-2966.2006.09996.x}{\emph{Mon. Not. R.
  Astron. Soc.} {\bfseries 367} (2006) 527}
  [\href{https://arxiv.org/abs/astro-ph/0507222}{{\ttfamily
  astro-ph/0507222}}].

\bibitem{bro+07}
J.~R. {Brownstein} and J.~W. {Moffat}, \emph{{The Bullet Cluster 1E0657-558
  evidence shows modified gravity in the absence of dark matter}},
  \href{https://doi.org/10.1111/j.1365-2966.2007.12275.x}{\emph{Mon. Not. R.
  Astron. Soc.} {\bfseries 382} (2007) 29}
  [\href{https://arxiv.org/abs/astro-ph/0702146}{{\ttfamily
  astro-ph/0702146}}].

\bibitem{mof+14}
J.~W. {Moffat} and S.~{Rahvar}, \emph{{The MOG weak field approximation - II.
  Observational test of Chandra X-ray clusters}},
  \href{https://doi.org/10.1093/mnras/stu855}{\emph{Mon. Not. R. Astron. Soc.}
  {\bfseries 441} (2014) 3724}
  [\href{https://arxiv.org/abs/1309.5077}{{\ttfamily 1309.5077}}].

\bibitem{mof+07}
J.~W. {Moffat} and V.~T. {Toth}, \emph{{Modified Gravity: Cosmology without
  dark matter or Einstein's cosmological constant}}, {\emph{ArXiv e-prints}
  (2007) } [\href{https://arxiv.org/abs/0710.0364}{{\ttfamily 0710.0364}}].

\bibitem{mof15c}
J.~W. {Moffat}, \emph{{Cosmological Evidence for Modified Gravity (MOG)}},
  {\emph{arXiv e-prints} (2015) arXiv:1510.07037}
  [\href{https://arxiv.org/abs/1510.07037}{{\ttfamily 1510.07037}}].

\bibitem{abb+17a}
B.~P. {Abbott}, R.~{Abbott}, T.~D. {Abbott}, F.~{Acernese}, K.~{Ackley},
  C.~{Adams} et~al., \emph{{GW170817: Observation of Gravitational Waves from a
  Binary Neutron Star Inspiral}},
  \href{https://doi.org/10.1103/PhysRevLett.119.161101}{\emph{Physical Review
  Letters} {\bfseries 119} (2017) 161101}
  [\href{https://arxiv.org/abs/1710.05832}{{\ttfamily 1710.05832}}].

\bibitem{abb+17b}
B.~P. {Abbott}, R.~{Abbott}, T.~D. {Abbott}, F.~{Acernese}, K.~{Ackley},
  C.~{Adams} et~al., \emph{{Gravitational Waves and Gamma-Rays from a Binary
  Neutron Star Merger: GW170817 and GRB 170817A}},
  \href{https://doi.org/10.3847/2041-8213/aa920c}{\emph{The Astrophysical
  Journal Letters} {\bfseries 848} (2017) L13}
  [\href{https://arxiv.org/abs/1710.05834}{{\ttfamily 1710.05834}}].

\bibitem{abb+17c}
B.~P. {Abbott}, R.~{Abbott}, T.~D. {Abbott}, F.~{Acernese}, K.~{Ackley},
  C.~{Adams} et~al., \emph{{Multi-messenger Observations of a Binary Neutron
  Star Merger}}, \href{https://doi.org/10.3847/2041-8213/aa91c9}{\emph{The
  Astrophysical Journal Letters} {\bfseries 848} (2017) L12}
  [\href{https://arxiv.org/abs/1710.05833}{{\ttfamily 1710.05833}}].

\bibitem{ezq+17}
J.~M. {Ezquiaga} and M.~{Zumalac{\'a}rregui}, \emph{{Dark Energy After
  GW170817: Dead Ends and the Road Ahead}},
  \href{https://doi.org/10.1103/PhysRevLett.119.251304}{\emph{Physical Review
  Letters} {\bfseries 119} (2017) 251304}
  [\href{https://arxiv.org/abs/1710.05901}{{\ttfamily 1710.05901}}].

\bibitem{par+18}
K.~{Pardo}, M.~{Fishbach}, D.~E. {Holz} and D.~N. {Spergel}, \emph{{Limits on
  the number of spacetime dimensions from GW170817}},
  \href{https://doi.org/10.1088/1475-7516/2018/07/048}{\emph{J. Cosmol.
  Astropart. Phys.} {\bfseries 7} (2018) 048}
  [\href{https://arxiv.org/abs/1801.08160}{{\ttfamily 1801.08160}}].

\bibitem{gre+18}
M.~A. {Green}, J.~W. {Moffat} and V.~T. {Toth}, \emph{{Modified gravity (MOG),
  the speed of gravitational radiation and the event GW170817/GRB170817A}},
  \href{https://doi.org/10.1016/j.physletb.2018.03.015}{\emph{Physics Letters
  B} {\bfseries 780} (2018) 300}
  [\href{https://arxiv.org/abs/1710.11177}{{\ttfamily 1710.11177}}].

\bibitem{per+17}
D.~{P{\'e}rez}, F.~G.~L. {Armengol} and G.~E. {Romero}, \emph{{Accretion disks
  around black holes in scalar-tensor-vector gravity}},
  \href{https://doi.org/10.1103/PhysRevD.95.104047}{\emph{Physical Review D}
  {\bfseries 95} (2017) 104047}
  [\href{https://arxiv.org/abs/1705.02713}{{\ttfamily 1705.02713}}].

\bibitem{guo+18}
M.~{Guo}, N.~A. {Obers} and H.~{Yan}, \emph{{Observational signatures of
  near-extremal Kerr-like black holes in a modified gravity theory at the Event
  Horizon Telescope}},
  \href{https://doi.org/10.1103/PhysRevD.98.084063}{\emph{Physical Review D}
  {\bfseries 98} (2018) 084063}
  [\href{https://arxiv.org/abs/1806.05249}{{\ttfamily 1806.05249}}].

\bibitem{won+18}
M.~F. {Wondrak}, P.~{Nicolini} and J.~W. {Moffat}, \emph{{Superradiance in
  modified gravity (MOG)}},
  \href{https://doi.org/10.1088/1475-7516/2018/12/021}{\emph{J. Cosmol.
  Astropart. Phys.} {\bfseries 12} (2018) 021}
  [\href{https://arxiv.org/abs/1809.07509}{{\ttfamily 1809.07509}}].

\bibitem{man+18}
L.~{Manfredi}, J.~{Mureika} and J.~{Moffat}, \emph{{Quasinormal modes of
  modified gravity (MOG) black holes}},
  \href{https://doi.org/10.1016/j.physletb.2017.11.006}{\emph{Physics Letters
  B} {\bfseries 779} (2018) 492}
  [\href{https://arxiv.org/abs/1711.03199}{{\ttfamily 1711.03199}}].

\bibitem{lop+17b}
F.~G. {Lopez Armengol} and G.~E. {Romero}, \emph{{Effects of
  Scalar-Tensor-Vector Gravity on relativistic jets}},
  \href{https://doi.org/10.1007/s10509-017-3197-6}{\emph{Astrophysics and Space
  Science} {\bfseries 362} (2017) 214}
  [\href{https://arxiv.org/abs/1611.09918}{{\ttfamily 1611.09918}}].

\bibitem{hus+15}
S.~{Hussain} and M.~{Jamil}, \emph{{Timelike geodesics of a modified gravity
  black hole immersed in an axially symmetric magnetic field}},
  \href{https://doi.org/10.1103/PhysRevD.92.043008}{\emph{Physical Review D}
  {\bfseries 92} (2015) 043008}
  [\href{https://arxiv.org/abs/1508.02123}{{\ttfamily 1508.02123}}].

\bibitem{mof15a}
J.~W. {Moffat}, \emph{{Black holes in modified gravity (MOG)}},
  \href{https://doi.org/10.1140/epjc/s10052-015-3405-x}{\emph{European Physical
  Journal C} {\bfseries 75} (2015) 175}
  [\href{https://arxiv.org/abs/1412.5424}{{\ttfamily 1412.5424}}].

\bibitem{lop+17a}
F.~G. {Lopez Armengol} and G.~E. {Romero}, \emph{{Neutron stars in
  Scalar-Tensor-Vector Gravity}},
  \href{https://doi.org/10.1007/s10714-017-2184-0}{\emph{General Relativity and
  Gravitation} {\bfseries 49} (2017) 27}
  [\href{https://arxiv.org/abs/1611.05721}{{\ttfamily 1611.05721}}].

\bibitem{ros15}
M.~{Roshan}, \emph{{Exact cosmological solutions for MOG}},
  \href{https://doi.org/10.1140/epjc/s10052-015-3637-9}{\emph{European Physical
  Journal C} {\bfseries 75} (2015) 405}
  [\href{https://arxiv.org/abs/1508.04243}{{\ttfamily 1508.04243}}].

\bibitem{jam+18}
S.~{Jamali}, M.~{Roshan} and L.~{Amendola}, \emph{{On the cosmology of
  scalar-tensor-vector gravity theory}},
  \href{https://doi.org/10.1088/1475-7516/2018/01/048}{\emph{Journal of
  Cosmology and Astroparticle Physics} {\bfseries 1} (2018) 048}
  [\href{https://arxiv.org/abs/1707.02841}{{\ttfamily 1707.02841}}].

\bibitem{mcv33}
G.~C. {McVittie}, \emph{{The mass-particle in an expanding universe}},
  \href{https://doi.org/10.1093/mnras/93.5.325}{\emph{Mon. Not. R. Astron.
  Soc.} {\bfseries 93} (1933) 325}.

\bibitem{far+07}
V.~{Faraoni} and A.~{Jacques}, \emph{{Cosmological expansion and local
  physics}}, \href{https://doi.org/10.1103/PhysRevD.76.063510}{\emph{Physical
  Review D} {\bfseries 76} (2007) 063510}
  [\href{https://arxiv.org/abs/0707.1350}{{\ttfamily 0707.1350}}].

\bibitem{car+10}
M.~{Carrera} and D.~{Giulini}, \emph{{Generalization of McVittie's model for an
  inhomogeneity in a cosmological spacetime}},
  \href{https://doi.org/10.1103/PhysRevD.81.043521}{\emph{Physical Review D}
  {\bfseries 81} (2010) 043521}
  [\href{https://arxiv.org/abs/0908.3101}{{\ttfamily 0908.3101}}].

\bibitem{tre+18}
D.~{Tretyakova} and B.~{Latosh}, \emph{{Scalar-Tensor Black Holes Embedded in
  an Expanding Universe}},
  \href{https://doi.org/10.3390/universe4020026}{\emph{Universe} {\bfseries 4}
  (2018) 26} [\href{https://arxiv.org/abs/1711.08285}{{\ttfamily 1711.08285}}].

\bibitem{mof+09}
J.~W. {Moffat} and V.~T. {Toth}, \emph{{Fundamental parameter-free solutions in
  modified gravity}},
  \href{https://doi.org/10.1088/0264-9381/26/8/085002}{\emph{Classical and
  Quantum Gravity} {\bfseries 26} (2009) 085002}
  [\href{https://arxiv.org/abs/0712.1796}{{\ttfamily 0712.1796}}].

\bibitem{rom13}
G.~E. {Romero}, \emph{{Adversus singularitates: The ontology of space-time
  singularities}}, {\emph{Foundations of Science} {\bfseries 18} (2013) 297}.

\bibitem{ell+77}
G.~F.~R. {Ellis} and B.~G. {Schmidt}, \emph{{Singular space-times}},
  \href{https://doi.org/10.1007/BF00759240}{\emph{General Relativity and
  Gravitation} {\bfseries 8} (1977) 915}.

\bibitem{poi04}
E.~{Poisson}, \emph{{A relativist's toolkit : the mathematics of black-hole
  mechanics}}. 2004.

\bibitem{far15}
V.~{Faraoni}, ed., \emph{{Cosmological and Black Hole Apparent Horizons}},
  vol.~907 of \emph{Lecture Notes in Physics, Berlin Springer Verlag}, 2015.
\newblock 10.1007/978-3-319-19240-6.

\bibitem{kal+10}
N.~{Kaloper}, M.~{Kleban} and D.~{Martin}, \emph{{McVittie's legacy: Black
  holes in an expanding universe}},
  \href{https://doi.org/10.1103/PhysRevD.81.104044}{\emph{Phys. Rev. D}
  {\bfseries 81} (2010) 104044}
  [\href{https://arxiv.org/abs/1003.4777}{{\ttfamily 1003.4777}}].

\bibitem{mof+08}
J.~W. {Moffat} and V.~T. {Toth}, \emph{{Testing Modified Gravity with Globular
  Cluster Velocity Dispersions}},
  \href{https://doi.org/10.1086/587926}{\emph{J. Cosmol. Astropart. Phys.}
  {\bfseries 680} (2008) 1158}
  [\href{https://arxiv.org/abs/0708.1935}{{\ttfamily 0708.1935}}].

\bibitem{pen69}
R.~{Penrose}, \emph{{Gravitational Collapse: the Role of General Relativity}},
  {\emph{Nuovo Cimento Rivista Serie} {\bfseries 1} (1969) }.

\end{thebibliography}\endgroup

\end{document}